# Line Identification and Excitation of Autoionizing States in a Late-Type, Low Mass Wolf-Rayet Star

Robert Williams[1,2], Catherine Manea[3], Bruce Margon[2], & Nidia Morrell[4]

[1] Space Telescope Science Institute, 3700 San Martin Drive, Baltimore, MD 21218, USA; wms@stsci.edu
[2] Department of Astronomy & Astrophysics, University of California, Santa Cruz, 1156 High Street, Santa Cruz, CA 95064, USA
[3] Department of Astronomy, University of Texas at Austin, Austin, TX 78712, USA
[4] Las Campanas Observatory, Carnegie Observatories, Casilla 601, La Serena, Chile

## Abstract

Identifications of a large fraction of previously unidentified lines in the complex spectrum of the low mass, late-type LMC [WC11] star J060819.93-715737.4 have been made utilizing electronic databases. There are an exceptionally large number of C II emission lines originating from autoionizing ("*ai*") levels. Resonance fluorescence between the C II ground state and excited *ai* levels is shown to be a important photoabsorption process that is competitive with dielectronic recombination in exciting *ai* emission lines in stellar winds, and has broad application to many types of emission-line stars. In addition, numerous C II quartet multiplets appear in emission that are not excited directly by recombination or resonance fluorescence, signifying high wind densities in the emission region that enhance collisional transfer between doublet and quartet states.

*Unified Astronomy Thesaurus concepts:* WC Stars; Stellar Spectral Lines

## 1. Introduction

The spectra of diffuse objects of low density, including planetary nebulae, old supernova remnants, and H II regions, are dominated by emission lines with strong forbidden transitions that differ significantly from those formed in the higher density emitting regions that characterize stellar winds. Much of the differences are due to the large disparity in density between the different classes of objects that dictates which types of radiative transitions can be prominent. However, a more important cause of the dissimilar spectra is the large difference in the dilution of the stellar continuum that drives the ionization and thermal balances in the objects. Fluorescence of stellar continuum radiation by line absorption plays at most a minor role in the population of excited levels in diffuse nebulae. However in stellar winds that originate near the photosphere it is a process that excites the emission-line spectrum.

The recent study of the late-type LMC [WC11] Wolf-Rayet star J060819.93-715737.4 (hereafter, J0608) by Margon et al. 2020a (hereafter, Paper I) illustrates what is involved in interpreting the visible spectrum of an object whose emission is formed in a stellar wind. The observed J0608 spectrum, shown in Figure 1, consists of emission and absorption lines formed from multiple ions. The emission lines are predominantly C II and He I transitions, whereas the absorption features are primarily due to Ne II and O II. Other late [WC] stars, including the prototype CPD -56° 8032 (De Marco et al. 1997), display similarly unusual spectra. The strongest emission lines in J0608 were identified in Paper I, but a more detailed attempt to identify a large number of weaker features, most of whom have not been cataloged previously for this type of object, was deferred. We report here on our more extensive effort to identify observed lines with specific transitions for which laboratory data exist and that may be considered astrophysically reasonable for a late type [WC] star with a prominent wind.

It should be noted that a number of the emission features observed in J0608 were identified in Paper I as originating from autoionizing (hereafter, "*ai*") levels above the ionization continuum, yet many of these transitions are not listed in authoritative databases such as the NIST Atomic Spectra Database[1] or CHIANTI[2]. Their identifications rely on a database making use of the important laboratory measurements of energy levels made by the Van de Graaff generator work by Bashkin & Stoner (1975, 1978). The existence of a multitude of transitions satisfying LS- and intermediate coupling rules has been gathered from these experimental data, yielding accurate wavelengths, and many such transitions have been listed in the v2.05 Atomic Line List[3] electronic database of van Hoof (2018). This database lists many transitions not found in other databases and serves as an important resource for the identification of lines from atomic species observed in astronomical spectra.

---

[1] https://www.nist.gov/pml/atomic-spectra-database

## 2. New Line Identifications

Following the publication of our initial results for J0608 we undertook a more thorough effort to identify transitions for which line IDs were not made in Paper I. Using the best signal/noise spectrum we have of J0608, observed with the Magellan MagE spectrograph on 2019 May 3, we re-measured line wavelengths for individual, relatively unblended features for which no satisfactory ID had been assigned in Paper I. Multiple databases were queried to find candidate transitions that were deemed reasonable based on multiple considerations. The usual factors of wavelength agreement, presence/absence of other members of the same multiplet, line width, ionization level, ion abundance, and oscillator strength when known were taken into consideration in assigning an identification. We have succeeded in making what we believe to be satisfactory line IDs for almost 150 additional lines that had not been assigned ID's in Paper I. All of the emission features which we consider to be real, based on their appearance in both the 2017 December 30 and 2019 May 3 spectra, are listed in Table 1. A small number of emission lines remain without satisfactory identification (ID) in spite of our efforts, and these are also listed.

Each of the lines in Table 1 is identified by its laboratory air wavelength to 0.1 Å and the spectroscopic terms of its lower and upper levels for its multiplet. Individual j-states are not denoted because in numerous cases individual multiplet members are not resolved in our spectrum. Thus, various emission features consist of multiple transitions within an unresolved multiplet. With the exception of some of the emission lines originating from *ai* levels, the listed wavelength and spectroscopic terms are sufficient to identify each transition specifically, virtually all of which appear in the NIST database. For emission lines originating from *ai* levels, we provide in Table 1 the electronic configurations of the lower and upper levels, which are sufficient to identify each transition----all of which appear in Bashkin & Stoner (1975) and the van Hoof (2018) v2.05 Line List. The upper and lower levels of all C II *ai* transitions identified in Table 1 have a $1s^2 2s2p(^3P^o)$ core term.

Several reasons factor into why the Identification of a given feature is unassigned. Lack of proper wavelength agreement is a primary consideration that introduces uncertainty in features that clearly consist of multiple transitions. Absence of a transition in the databases is also a factor, normally resulting from either the lack of experimental data or inadequate theoretical calculations characterizing the transition. This is particularly true for *ai* levels more than a few eV above the ionization continuum. In some cases there are multiple transitions that are reasonable IDs and it is difficult to determine which of them are the proper ID, even after taking into account the presence or absence of associated multiplet members.

Several of the IDs in Table 1 are worthy of comment, especially the transitions generally associated with low density nebular gas. The forbidden [O II] λ3727 doublet is prominent, as are the [S II] λ6717/34, [N II] 6548/84, and [O I] λ6300/63 doublets that are not expected to arise in a stellar wind. As noted in Paper 1, the known Galactic late-type [WC] stars are surrounded by PNe. Although none has been detected in images of J0608 the presence of the above nebular lines in its spectrum does indicate the likely existence of what could be a low surface brightness PN ejected by the object at an earlier epoch.  Margon et al. (2020b) have noted that at the LMC distance, a PN of physical extent and surface brightness similar to that seen in the prototype Galactic [WC11] star CPD -56º 8032 would not be resolved in ground-based images.

---
[2] https://www.chiantidatabase.org/chianti.html
[3] http://www.pa.uky.edu/~peter/newpage/

  The subjective judgment of what one considers astrophysical reasonableness certainly introduces its own bias.  It is possible that some of the unassigned transitions in Tables 1 of this paper and Paper II are indeed listed in databases but lacking information on expected associated lines, thus resulting in uncertainty for an otherwise valid ID so that it is rejected.  Complete assurance of the identification of a specific transition is often not possible, and self-consistency with other line IDs is often the strongest argument for a particular ID.

### 3.  Excitation of Autoionization States

  The interesting mixture of numerous C II emission lines in the J0608 spectrum with Ne II and O II absorption lines was described in Paper I.  One feature of the spectrum discussed in that paper is the significance of multiple strong C II emission lines originating from autoionizing levels.  De Marco, Storey, & Barlow (1997) had previously called attention to the presence of such transitions in the low-mass [WC10] stars CPD-56º 8032 and He 2-113, and they considered the different processes that produce the emission from *ai* levels.  *Ai* transitions were also noted in the spectrum of V348 Sgr (Dahari & Osterbrock 1984; Leuenhagen, Heber, & Jeffery 1994), which although classified as a hot R CrB star (Crowther, De Marco, & Barlow 1998), does share characteristics of the late [WC] class.  The presence of numerous transitions between *ai* levels in J0608 does suggest that they may be a characteristic of winds with enhanced carbon abundance.

  The mechanism that populates *ai* levels is generally acknowledged to be dielectronic recombination (DR) (Seaton & Storey 1976; Nussbaumer & Storey 1984; Pradhan & Nahar 2011).  Using the CPD-56º 8032 spectrum, De Marco et al. (1997) performed a detailed study of its formation and the relative intensities of the emission lines.  Their analysis, which included *ai* doublet transitions, used fits to the profiles of individual members of *ai* multiplets to determine relative optical depths and accurate line fluxes for the lines.  They found C II excited bound-bound transitions, including *ai* doublets and strong radiative recombination lines like $\lambda$4267, to be optically thick in the CPD-56º 8032 wind.  Their model, which neglected absorption of the stellar continuum by lines, did demonstrate that many line optical depths can sufficiently large that this needs to be accounted for when determining physical conditions and element abundances.

  *Ai* levels are doubly excited states and photoexcitation of an ion in the ground state does not normally produce excitation to an *ai* level because the cross sections for radiative excitation are larger for both singly excited bound states and for photoionization.  There are exceptions, however.  Absorption of a photon with ionization energy by an electron in the outer shell, but of the inner *sub*-shell of the valence shell, can result in excitation of the inner *sub*-shell electron to a bound *ai* level.  This produces a doubly excited *ai* state, and it is possible for the absorption to occur via electric dipole transitions obeying LS-coupling selection rules with high transition probabilities.  Such absorptions, which are frequently followed by autoionization, appear as prominent resonances in the photoionization cross sections of ions (Pradhan & Nahar 2011).

  The above photoexcitation mechanism was suggested in Paper I as a possible cause for the numerous strong *ai* transitions in the J0608 spectrum.  Their intensities originating from both C II doublet and quartet *ai* levels were shown to be more than an order of magnitude greater than can be accounted for by dielectronic recombination alone, assuming the lines to be optically thin.  This is especially significant in understanding the population of C II quartet *ai* levels, which are not populated directly by DR from the C III singlet ground state via LS-coupling, thus suggesting collisional transfer between the doublet and quartet states. However, because the results of De Marco et al. (1997) have shown that the emission lines may not be optically thin as we had assumed for simplicity, this matter needs to be pursued further.

Three processes can populate *ai* levels by the absorption of continuum radiation: (1) resonance-line absorption from the ground state to *ai* levels; (2) absorption from excited levels populated by optically thick resonance-line scattering, from which transitions to *ai* levels with large f-values exist (Dahari & Osterbrock 1984); and (3) a stellar continuum that is sufficiently strong to produce adequate photoexcitation. Absorption of stellar continuum radiation is not usually considered a significant process, however there are conditions where it could be relevant for C II in late [WC] stars like J0608 and contribute to observed *ai* line strengths.

The rate $R_j$ at which DR populates level j of ion i is given by

$$R_j = n_e\, n_{i+1}\, \alpha_{DR}(j) \qquad cm^{-3}\, s^{-1} \tag{1}$$

where $n_e$ and $n_{i+1}$ are the electron and ion number densities, and $\alpha_{DR}(j)$ is the dielectronic recombination coefficient for level j. DR coefficients have been calculated for numerous ions by Storey (1981), Nussbaumer & Storey (1984), and Sochi & Storey (2013), and are sensitive to electron temperature $T_e$ for *ai* levels with excitation potentials significantly above the ionization limit of ion i. However, they are insensitive to $T_e$ for levels whose excitation potentials are comparable to or less than the mean free electron energy.

The rate $P_{1j}$ at which photoabsorptions from the ground state populate level j of ion i is

$$P_{1j} = n_i\, B_{1j}\, J_{\nu 1j} \qquad cm^{-3}\, s^{-1} \tag{2}$$

where $B_{1j}$ is the Einstein stimulated emission coefficient for the transition and $J_{\nu 1j}$ the mean intensity of radiation from the star+wind continuum at the resonance line frequency $\nu_{1j}$. The relative abundance of $C^+$ and $C^{+2}$ ions is determined by the ionization equation

$$n_{C+} \int_{\nu_c}^{\infty} 4\pi J_\nu\, a_\nu \frac{d\nu}{h\nu} = n_e\, n_{C+2}\, \alpha^{rec}(T_e) \tag{3}$$

where $\nu_c$, $a_\nu$, and $\alpha^{rec}(T_e)$ are the ionization frequency threshold, photoionization cross section, and total radiative recombination coefficient, respectively, for $C^+$.

For the wavelength regime relevant for the carbon ions present in the J0608 spectrum we represent the continuum radiation field by the Wien approximation due to J0608's relatively low photospheric temperature of $T_* \leq 30{,}000$ K, estimated from the spectral energy distribution, which causes the maximum black body radiation to remain longward of 1,000 Å. Given the steep decline in the continuum flux at higher frequencies the photoionization cross section for C II can be approximated well by $a_\nu = a_c \times (\nu_c/\nu)^2$, with $a_c = 4.6 \times 10^{-18}$ cm$^2$ (Nahar & Pradhan 1997). Taking $J_\nu = J_c \times (\nu/\nu_c)^3 \exp(-h\nu/kT_*)$, eqn (3) leads to

$$n_{C+} = n_e\, n_{C+2} \times h^2 \nu_c\, \alpha^{rec}(T_e) / (4\pi J_c\, a_c\, kT_*) \times \exp(h\nu_c/kT_*). \tag{4}$$

Using the relationship between the transition oscillator strength and Einstein coefficient, $B_{1j} = 4\pi^2 e^2 f_{1j}/(m_e c h \nu_{1j})$ (Mihalas 1978), the rate of excitations to a level j by photoexcitation from the C II ground state, $P_{1j}$, becomes

$$P_{1j} = n_{C+} \times 4\pi^2 e^2 f_{1j}/(m_e c h \nu_{1j})\, J_c\, (\nu_{1j}/\nu_c)^3 \exp(-h\nu_{1j}/kT_*). \tag{5}$$

A comparison of the relative rate of photoexcitation to that of DR in populating a level j is given by the ratio of the two rates,

$$P_{1j}/R_j = \pi h e^2 f_{1j}/(m_e c a_c k T_*) \times (\lambda_c/\lambda_{1j})^2 \times \alpha^{rec}(T_e)/\alpha_{DR}(j) \times \exp\{h(\nu_c-\nu_{1j})/kT_*\}. \quad (6)$$

This relation can be applied to the excitation of the C II $2s2p(^3P^o)4p$ $^2P$ state from both radiative and dielectronic recombination, as has been considered by Nahar (1995), and continuum photoexcitation via the $2s^22p$ $^2P^o - 2s2p4p$ $^2P$ $\lambda 466$ Å UV resonance line to the upper autoionization level. The DR coefficient for a level can be derived from the published emission coefficient by correcting for the autoionization probability for the level and the branching ratio for the transition. However, rather than determine the DR rate for individual C II $ai$ levels we consider the more appropriate population rate for all $ai$ levels collectively, $R_{\Sigma j}$, using the value of the DR rate for $ai$ levels that has been calculated by Davey et al. (2000), $\alpha_{DR}(C^+) = 5\times10^{-12}$ cm$^3$ s$^{-1}$.

In determining the rate of photoexcitation from the ground state to the single $ai$ level $2p(^3P^o)4p$ $^2P$ due to the $\lambda 466$ Å resonance line, the photospheric temperature is taken to be $T_* = 28{,}000$ K, the approximate determined value from Paper 1. We also select the electron temperature $T_e = 10^4$ K for the calculation, with the atomic parameters $\alpha^{rec}(C^+) = 6.0\times10^{-12}$ cm$^3$ s$^{-1}$ (Nahar 1995), $f_{\lambda 466} = 0.018$ (Kramida et al. 2019), and $\lambda_c = 508$ Å. Using these values the ratio of the continuum photoexcitation rate for the one resonance line to its upper $ai$ level relative to the DR rate for all $ai$ levels is $P_{1j}/R_{\Sigma j} = 0.10$. When additional resonance lines are considered, not to mention a higher value of $T_*$ that would provide more UV photons, resonance-line absorption of continuum radiation is definitely a competitive process with dielectronic recombination in populating autoionization states. Similar values of the contribution of continuum absorption from excited bound to $ai$ levels yield similar values of $P_{ij}/R_j$. These figures demonstrate that the resonance fluorescence process should be taken into account when considering the excitation of $ai$ transitions.

The question of how line optical depths influence the population of excited states remains to be understood, especially in relation to the excited levels that do not couple directly to the ground state via strong resonance lines, i.e., those that are allowed by LS-coupling. In the case of C II this is especially important for the quartet levels that are not reached directly by radiative or dielectronic recombination from the C III singlet ground state nor by resonance-line scattering from the C II doublet ground state

As is evident from Table 1, C II quartet emission lines that include a number originating from $ai$ levels do appear with appreciable strength in J0608. For wind densities greater than $10^{12}$ cm$^{-3}$ collisional transfer between excited levels of every multiplicity that have similar excitation potentials will be a dominant process, just as it is for the O I triplets and quintets (Bhatia & Kastner 1995; Kastner & Bhatia 1995). It is very likely to be responsible for the population of many of the C II quartet states in J0608. Construction of a wind model for J0608 that considers known radiative and collisional processes is surely one of the best ways to determine the significance of the different processes that produce the object's spectrum. Based on the calculations above, in the future such models should take into account the fact that ground-state absorption is a viable process that must be considered in the population of $ai$ levels.

## 4. Summary

The large majority of emission lines that appear in the spectrum of the [WC11] star J0608 have now been identified from information obtained in the NIST and V2.05 Atomic Line Lists. The relatively narrow line widths of ~90 km/s FWHM for this type of object have enabled most multiplets to be resolved, aiding in their identification. However, the large number of candidate transitions that occur within the acceptable wavelength interval for most of the observed lines does cause sufficient uncertainty that inevitable errors in identification will occur.

The emission spectrum contains an unusually large number of transitions originating from C II autoionizing levels.  We have shown that resonance fluorescence from the ground state can play a prominent role in populating *ai* levels.  Uncertainties in the J0608 wind properties do not allow a more definite statement to be made as to which processes dominate the excitation of a given level without the aid of a model that treats radiative transfer in the wind.  Such models do exist, e.g., CMFGEN developed by D. J. Hillier and colleagues (Hillier & Miller 1998; Hillier & Lanz 2001) is one example, and would be very useful in confirming those physical processes that play a role in the formation of spectra of late [WC] and R CrB stars.

The authors thank our colleague J. Xavier Prochaska for initially calling this object to our attention, and John Hillier for many helpful conversations on line formation in winds.  The Space Telescope Science Institute is operated by the Associations of Universities for Research in Astronomy, Incorporated, under NASA contract NAS5-26555.

# Appendix A: Possible Rydberg Enhanced Recombination

A recent study by Nemer et al. (2019) has called attention to a process called Rydberg Enhanced Recombination (RER) that can be important in populating bound states just below the ionization threshold. It enhances the emission by radiative decay from highly excited bound levels but has little effect in augmenting excitation of *ai* levels. There is a test to determine the relative importance of cascades from *ai* levels vs. the RER process in populating bound levels near the ionization limit. It is a comparison of the relative intensities of the 'feed' transitions from *ai* levels into a highly bound level compared to the intensities of the emission lines out of that level.

The high bound level of C II 2s2p3d $^4$F$^o$ is the upper level of the $\lambda$7118 $^4$D-$^4$F$^o$ multiplet transition. Because of its close proximity within 0.1 eV of the C$^+$ ionization limit that level is a good candidate for enhanced population by RER. The level is also populated radiatively by cascades from the C II $\lambda$3877 and $\lambda$5258 multiplets, whose upper levels are *ai* states that directly feed the $\lambda$7118 emission feature. When the $\lambda$3877 + $\lambda$5258 multiplet fluxes are similar to or exceed the flux of $\lambda$7118, it indicates that cascades from the *ai* levels may be the predominant source of excitation of the 3d $^4$F$^o$ level. Our measurements from the 3 May 2019 MagE spectrum yield an observed flux ratio of F($\lambda$3877+$\lambda$5258)/F($\lambda$7118) ~ 2.3, not accounting for probable modest reddening that increases the ratio. Thus, radiative cascading from *ai* levels into the $^4$F$^o$ upper level of $\lambda$7118 should lead to it having a higher intensity than observed unless a non-radiative de-excitation of that level is also taking place, e.g., collisional transfer to other states. Under these circumstances, there is no substantive evidence that RER contributes to the population of the 3d $^4$F$^o$ state in J0608.

**Figure 1**

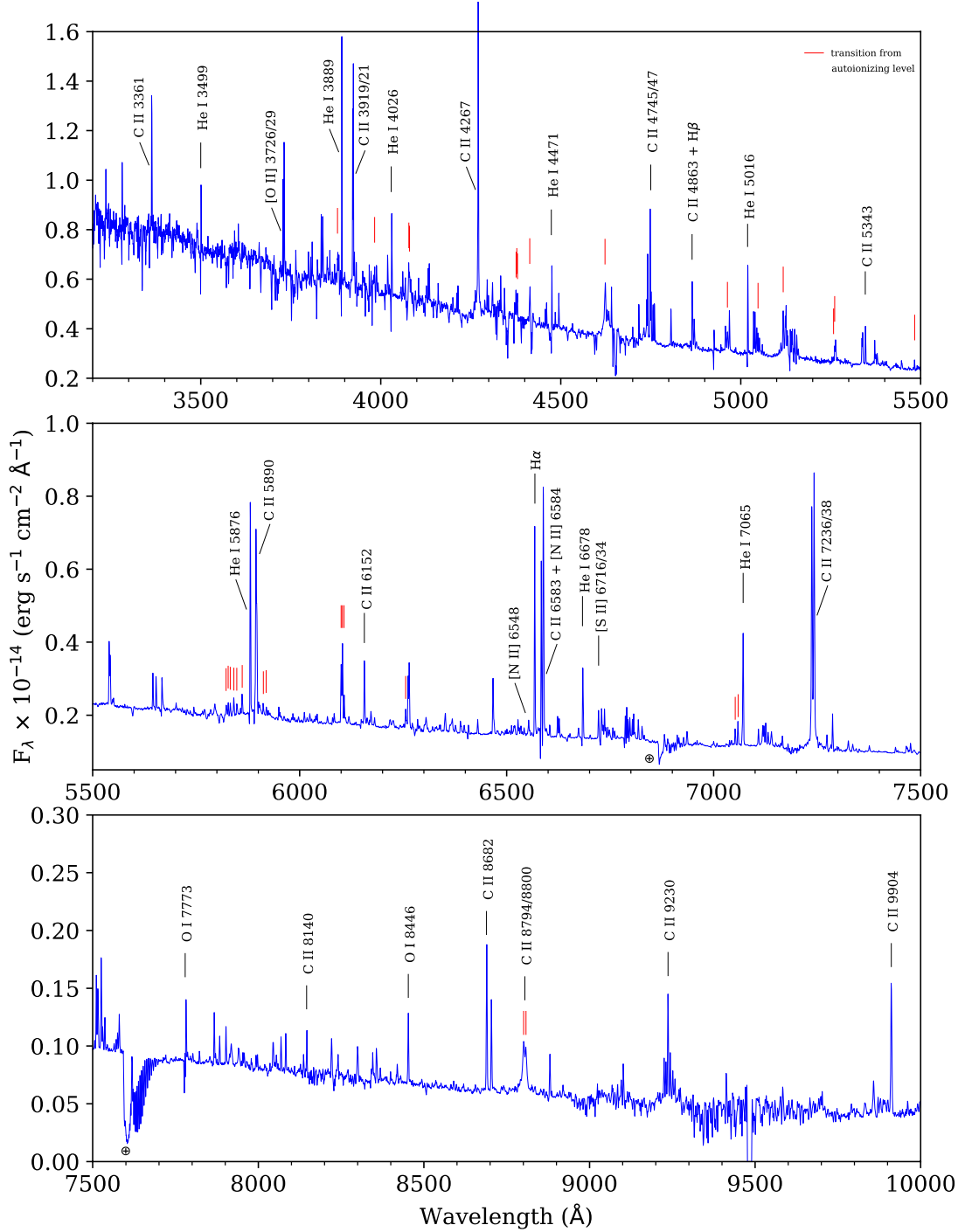

**Figure 1** - Spectrum of the LMC [WC11] star J0608, taken with the Magellan MagE spectrograph on 2019 May 3. The large number of transitions originating from autoionizing levels is a unique feature of this spectrum, as are the large number of C II emission lines, listed in Table 1, that completely dominate the spectrum. The forbidden lines almost certainly originate in a surrounding nebula that is spatially unresolved.

Table 1. Identification and Measurements of J0608 Spectral Lines

| Line ID (Å) | Transition | Type | $\lambda_{observed}$ (Å) |
|---|---|---|---|
| O II 3134.2 | $^4D^o$–$^4P$ | abs | 3137.1 |
| O II 3138.3 | $^4D^o$–$^4P$ | abs | 3141.7 |
| Fe I 3152.0 | $^5F^o$–$^5F$ | abs | 3154.9 |
| C II 3165.5 | $^2D^o$–$^2P$ | em | 3168.0 |
| C II 3167.9 | $^2D^o$–$^2P$ | em | 3170.3 |
| He I 3187.7 | $^3S$–$^3P^o$ | P Cyg | 3189.9 |
| Ne II 3213.7 | $^4D^o$–$^4F$ | abs | 3216.4 |
| Ne II 3217.3 + O II 3217.9 | $^4D^o$–$^4P$ + $^6P$–$^6S^o$ | abs | 3220.4 |
| Ne II 3230.1 | $^2D$–$^2D^o$ | abs | 3232.4 |
| Ne II 3232.3 | $^2D$–$^2D^o$ | abs | 3236.4 |
| ? | | P Cyg | 3281.1 |
| Ne II 3297.7 | $^4P$–$^4D^o$ | abs | 3300.3 |
| Ne II 3323.7 | $^2P$–$^2P^o$ | abs | 3326.1 |
| Ne II 3327.2 | $^4P$–$^4D^o$ | abs | 3329.8 |
| Ne II 3334.8 | $^4P$–$^4D^o$ | abs | 3337.3 |
| Ne II 3344.4 | $^4P$–$^4D^o$ | abs | 3347.4 |
| Ne II 3355.0 | $^4P$–$^4D^o$ | abs | 3357.6 |
| C II 3361.3 | $^2D$–$^2P^o$ | em | 3364.0 |
| Ne II 3367.2 | $^2D^o$–$^2F$ | abs | 3370.3 |
| Ne II 3378.2 | $^2P$–$^2P^o$ | abs | 3380.4 |
| Ne II 3388.4 | $^2D^o$–$^2F$ | abs | 3391.2 |
| Ne II 3392.8 | $^2P$–$^2P^o$ | abs | 3395.6 |
| Ne II 3416.9 | $^2D^o$–$^2D$ | abs | 3420.2 |
| He I 3447.3 | $^1S$–$^1P^o$ | abs | 3449.8 |
| O II 3470.7 | $^2D^o$–$^2P$ | abs | 3473.3 |
| Ne II 3481.9 | $^2P$–$^2S^o$ | abs | 3484.3 |
| He I 3498.6 | $^3P^o$–$^3D$ | P Cyg | 3500.3 |
| He I 3512.5 | $^3P^o$–$^3D$ | abs | 3514.7 |
| Ne II 3543.9 | $^4S^o$–$^4P$ | abs | 3545.3 |
| Ne II 3568.5 | $^2D$–$^2F^o$ | abs | 3571.2 |
| Ne II 3574.6 | $^2D$–$^2F^o$ | abs | 3577.3 |
| He I 3587.3 | $^3P^o$–$^3D$ | abs | 3589.6 |
| C II 3590.8 | $^4D$–$^4P^o$ | em | 3593.6 |
| ? | | em | 3604.5 |
| C III 3609.3 | $^3P^o$–$^3D$ | abs | 3611.9 |
| He I 3613.6 | $^1S$–$^1P^o$ | abs | 3616.0 |
| He I 3634.3 | $^3P^o$–$^3D$ | abs | 3636.5 |

**Table 1** *(continued)*

| Line ID (Å) | Transition | Type | $\lambda_{observed}$ (Å) |
|---|---|---|---|
| Ne II 3664.1 | $^4P$–$^4P^o$ | abs | 3666.7 |
| Ne II 3694.2 | $^4P$–$^4P^o$ | abs | 3696.9 |
| C II 3700.4 | $^2D$–$^2F^o$ | em | 3703.6 |
| He I 3704.7 + 3705.0 | $^1P^o$–$^1D$ + $^3P^o$–$^3D$ | abs | 3707.4 |
| Ne II 3709.6 | $^4P$–$^4P^o$ | abs | 3712.5 |
| O II 3712.7 | $^4P$-$^4S^o$ | abs | 3715.4 |
| [O II] 3726.0 | $^4S^o$–$^2D^o$ | em | 3728.7 |
| [O II] 3728.8 | $^4S^o$–$^2D^o$ | em | 3731.5 |
| H I 3734.3 | 2–13 | abs | 3737.8 |
| O II 3749.5 | $^4P$–$^4S^o$ | abs | 3752.1 |
| Ne II 3766.3 | $^4P$–$^4P^o$ | abs | 3769.1 |
| C II 3796.8 | $^2F^o$–$^2G$ | em | 3799.2 |
| C II 3802.1 | $^2D$–$^2F^o$ | em | 3805.4 |
| He I 3805.7 + C III 3806.4 | $^3P^o$–$^3D$ | em | 3809.5 |
| He I 3818.9 | $^3P^o$–$^3D$ | abs | 3822.0 |
| C II 3831.7 | $^2P^o$–$^2D$ | em | 3834.6 |
| C II 3835.7 + H9 | $^2P^o$–$^2D$ | em | 3839.0 |
| ? | | em | 3851.1 |
| Si II 3856.0 | $^2D$–$^2P^o$ | em | 3858.5 |
| C II 3876.4 | 2p3d $^4F^o$–2p4f $^4G$ | em | 3880.5 |
| He I 3888.6 | $^3S$–$^3P^o$ | P Cyg | 3891.3 |
| C II 3919.0 | $^2P^o$–$^2S$ | em | 3922.0 |
| C II 3920.7 | $^2P^o$–$^2S$ | em | 3923.8 |
| C II 3923.8 | $^2D$–$^2F^o$ | em | 3926.9 |
| C II 3929.0 + C II 3928.8 | $^2F^o$–$^2G$ + $^2G$-$^2F^o$ | em | 3931.9 |
| O II 3954.4 | $^2P$–$^2P^o$ | abs | 3957.2 |
| He I 3964.7 | $^1S$–$^1P^o$ | P Cyg | 3967.4 |
| Hε 3970.1 | 2–7 | em | 3973.1 |
| O II 3973.3 | $^2P$–$^2P^o$ | abs | 3976.2 |
| C II 3980.3 | 2p3d $^4D^o$–2p4f $^4D$ | em | 3983.3 |
| C II 3985.1 | $^2F^o$–$^2G$ | em | 3988.3 |
| He I 4009.3 | $^1P^o$–$^1D$ | P Cyg | 4012.3 |
| C II 4015.7 | $^2D$–$^2F^o$ | em | 4019.0 |
| He I 4026.2 | $^3P^o$–$^3D$ | P Cyg | 4029.4 |
| O II 4069.6 | $^4D^o$–$^4F$ | abs | 4072.8 |
| C II 4075.5 | 2p3d $^4D^o$–2p4f $^4F$ | em | 4077.7 |
| C II 4077.6 | 2p3d $^4D^o$–2p4f $^4F$ | em | 4082.0 |
| Hδ 4101.7 | 2–6 | em | 4104.9 |
| He I 4120.8 | $^3P^o$–$^3S$ | abs | 4123.0 |

*Table 1 continued on next page*

| Line ID | Transition | Type | $\lambda_{observed}$ |
| --- | --- | --- | --- |
| (Å) | | | (Å) |
| ? | | em | 4126.8 |
| Si II 4128.1 | $^2D$–$^2F^o$ | em | 4131.8 |
| Si II 4130.9 | $^2D$–$^2F^o$ | em | 4134.9 |
| He I 4143.8 | $^1P^o$–$^1D$ | abs | 4146.0 |
| Ne II 4152.4 | $^2P$–$^2P^o$ | abs | 4155.6 |
| C II 4156.1 | $^2F^o$–$^2G$ | em | 4159.4 |
| O I 4207.3 | $^3D$–$^3P^o$ | em | 4211.1 |
| O I 4211.7 | $^3D$–$^3P^o$ | em | 4214.5 |
| Ne II 4222.1 | $^2P$–$^2P^o$ | abs | 4224.9 |
| C II 4267.3 | $^2D$–$^2F^o$ | em | 4270.7 |
| C II 4292.3 | $^2F^o$–$^2G$ | em | 4295.6 |
| C II 4306.3 | $^2P^o$–$^2S$ | em | 4310.1 |
| O II 4317.1 | $^4P$–$^4P^o$ | abs | 4320.1 |
| O II 4319.6 | $^4P$–$^4P^o$ | abs | 4322.9 |
| C II 4329.8 | $^2D$–$^2F^o$ | em | 4333.2 |
| O II 4336.9 | $^4P$–$^4P^o$ | abs | 4340.0 |
| H$_\gamma$ | | em | 4343.8 |
| O II 4345.6 | $^4P$–$^4P^o$ | abs | 4348.9 |
| O II 4349.4 | $^4P$–$^4P^o$ | abs | 4352.6 |
| O II 4366.9 | $^4P$–$^4P^o$ | abs | 4370.0 |
| C II 4372.4 | 2p3d $^4P^o$–2p4f $^4D$ | em | 4375.6 |
| C II 4374.3 | 2p3d $^4P^o$–2p4f $^4D$ | em | 4378.3 |
| C II 4376.6 | 2p3d $^4P^o$–2p4f $^4D$ | em | 4379.7 |
| He I 4387.9 | $^1P^o$–$^1D$ | abs | 4390.6 |
| C II 4411.3 | 2p3d $^2D^o$–2p4f $^2F$ | em | 4414.2 |
| O II 4414.9 | $^2P$–$^2D^o$ | abs | 4418.3 |
| O II 4417.0 | $^2P$–$^2D^o$ | abs | 4420.4 |
| ? | | em | 4457.0 |
| ? | | em | 4460.5 |
| He I 4471.5 | $^3P^o$–$^3D$ | P Cyg | 4474.5 |
| C II 4491.2 | $^2F^o$–$^2G$ | em | 4494.7 |
| C III 4516.3 | $^3P^o$-3S | abs | 4519.6 |
| O II 4591.0 | $^2D$–$^2F^o$ | abs | 4594.4 |
| O II 4596.2 | $^2D$–$^2F^o$ | abs | 4599.4 |
| C II 4619.2 | 2p3d $^2F^o$–2p4f $^2G$ | em | 4623.2 |
| C II 4637.6 | $^2P^o$–$^2D$ | em | 4640.9 |
| O II 4641.8 | $^4P$–$^4D^o$ | abs | 4645.1 |
| O II 4649.1 | $^4P$–$^4D^o$ | abs | 4652.6 |
| O II 4650.8 | $^4P$–$^4D^o$ | abs | 4654.4 |

**Table 1** *(continued)*

| Line ID | Transition | Type | $\lambda_{observed}$ |
|---|---|---|---|
| (Å) | | | (Å) |
| O II 4661.6 | $^4P$–$^4P^o$ | abs | 4664.9 |
| O II 4676.2 | $^4P$–$^4D^o$ | abs | 4679.6 |
| O I 4696.5 | $^3D$–$^3P^o$ | em | 4699.9 |
| He I 4713.2 | $^3P^o$–$^3S$ | P Cyg | 4716.4 |
| C II 4735.5 | $^2P$–$^2P^o$ | em | 4739.1 |
| C II 4738.0 | $^2P$–$^2P^o$ | em | 4741.6 |
| C II 4744.8 | $^2P$–$^2P^o$ | em | 4748.5 |
| C II 4747.3 | $^2P$–$^2P^o$ | em | 4750.8 |
| ? | | em | 4756.5 |
| ? | | em | 4760.3 |
| C II 4802.7 | $^2F^o$–$^2G$ | em | 4806.2 |
| C II 4862.6 + H,8 | $^2S$–$^2P^0$ | em | 4865.5 |
| C II 4867.1 | $^2S$–$^2P^o$ | em | 4870.6 |
| He I 4921.9 | $^1P^o$–$^1D$ | P Cyg | 4925.3 |
| ? | | P Cyg | 4944.7 |
| C II 4953.9 | 2p3p $^2P$–2p3d $^2P^o$ | em | 4957.5 |
| C II 4958.7 | 2p3p $^2P$–2p3d $^2P^o$ | em | 4962.9 |
| C II 4964.7 | 2p3p $^2P$–2p3d $^2P^o$ | em | 4968.5 |
| He I 5015.7 | $^1S$–$^1P^o$ | P Cyg | 5019.0 |
| C II 5032.1 | $^2P^o$–$^2D$ | em | 5035.8 |
| C II 5035.9 | $^2P^o$–$^2D$ | em | 5039.6 |
| C II 5040.7 | $^2P^o$–$^2D$ | em | 5044.6 |
| C II 5044.4 | 2p3d $^4D^o$–2p4p $^4P$ | em | 5048.3 |
| He I 5047.7 | $^1P^o$–$^1S$ | em | 5052.0 |
| C II 5113.9 | 2p3d $^2P^o$–2p4f $^2D$ | em | 5117.9 |
| C II 5120.1 | $^2P^o$–$^2P$ | em | 5123.5 |
| C II 5121.8 | $^2P^o$–$^2P$ | em | 5125.7 |
| C II 5125.2 | $^2P^o$–$^2P$ | em | 5129.1 |
| C II 5132.9 | $^4P^o$–$^4P$ | P Cyg | 5136.2 |
| C II 5133.3 | $^4P^o$–$^4P$ | em | 5137.8 |
| C II 5137.3 | $^4P^o$–$^4P$ | em | 5141.4 |
| C II 5139.2 | $^4P^o$–$^4P$ | em | 5143.4 |
| C II 5143.5 | $^4P^o$–$^4P$ | P Cyg | 5146.7 |
| C II 5145.2 | $^4P^o$–$^4P$ | P Cyg | 5149.1 |
| C II 5151.1 | $^4P^o$–$^4P$ | P Cyg | 5154.9 |
| C II 5257.2 | 2p3d $^4F^o$–2p4p $^4D$ | em | 5261.1 |
| C II 5259.5 | 2p3d $^4F^o$–2p4p $^4D$ | em | 5263.5 |
| C II 5332.9 | $^2P^o$–$^2S$ | em | 5336.6 |
| C II 5334.8 | $^2P^o$–$^2S$ | em | 5338.5 |

**Table 1** *(continued)*

| Line ID | Transition | Type | $\lambda_{observed}$ |
|---|---|---|---|
| (Å) | | | (Å) |
| C II 5342.4 | $^2F^o$–$^2G$ | em | 5346.4 |
| C II 5368.6 | $^2D$–$^2P^o$ | em | 5372.3 |
| C II 5370.4 | $^4P^o$–$^4D$ | em | 5374.6 |
| C II 5374.2 | $^4P^o$–$^4D$ | em | 5378.9 |
| C IV 5411.4 + He II 5411.5 | $^2G$–$^2H^o$ | abs | 5415.9 |
| C II 5478.6 | 2p3d $^4D^o$–2p4p $^4D$ | em | 5482.7 |
| O III 5508.2 | $^1D$–$^1D^o$ | em | 5512.6 |
| C II 5535.4 | $^2S$–$^2P^o$ | em | 5539.5 |
| C II 5537.6 | $^2S$–$^2P^o$ | em | 5541.8 |
| O III 5592.3 | $^1P^o$–$^1P$ | abs | 5596.3 |
| C II 5640.5 | $^4P^o$–$^4S$ | em | 5645.2 |
| C II 5648.1 | $^4P^o$–$^4S$ | P Cyg | 5651.9 |
| C II 5662.5 | $^4P^o$–$^4S$ | P Cyg | 5666.5 |
| C III 5695.9 | $^1P^o$–$^1D$ | em | 5701.3 |
| ? | | P Cyg | 5744.2 |
| C IV 5801.4 | $^2S$–$^2P^o$ | abs | 5805.5 |
| C IV 5812.0 | $^2S$–$^2P^o$ | abs | 5816.2 |
| C II 5818.3 | 2p3p $^4D$–2p3d $^4P^o$ | em | 5822.4 |
| C II 5823.2 | 2p3p $^4D$–2p3d $^4P^o$ | em | 5827.4 |
| C II 5827.9 | 2p3p $^4D$–2p3d $^4P^o$ | em | 5832.2 |
| C II 5836.4 | 2p3p $^4D$–2p3d $^4P^o$ | em | 5840.6 |
| C II 5843.6 | 2p3p $^4D$–2p3d $^4P^o$ | em | 5848.4 |
| C II 5856.1 | 2p3p $^4D$–2p3d $^4P^o$ | em | 5860.5 |
| He I 5875.6 | $^3P^o$–$^3D$ | P Cyg | 5879.4 |
| C II 5889.8 | $^2D$–$^2P^o$ | em | 5894.8 |
| C II 5907.2 | 2p3d $^4P^o$–2p4p $^4S$ | em | 5911.6 |
| C II 5914.6 | 2p3d $^4P^o$–2p4p $^4S$ | em | 5919.3 |
| C II 6006.0 | $^2D$–$^2F^o$ | em | 6011.2 |
| ? | | em | 6024.3 |
| C II 6034.2 | $^4D^o$–$^4F$ | em | 6038.6 |
| C II 6037.8 | $^4D^o$–$^4F$ | em | 6041.9 |
| C II 6061.8 | $^2F^o$–$^2G$ | em | 6067.0 |
| C II 6077.9 | $^2D$–$^2F^o$ | em | 6083.2 |
| C II 6095.3 | 2p3p $^2P$–2p3d $^2D^o$ | em | 6100.0 |
| C II 6098.5 | 2p3p $^2P$–2p3d $^2D^o$ | em | 6103.1 |
| C II 6102.6 | 2p3p $^2P$–2p3d $^2D^o$ | em | 6107.4 |
| C II 6113.4 | $^2F^o$–$^2G$ | em | 6118.7 |
| C II 6151.5 | $^2D$–$^2F^o$ | em | 6156.2 |
| C II 6167.0 | $^2D$–$^2F^o$ | em | 6171.9 |

**Table 1** *(continued)*

| Line ID (Å) | Transition | Type | $\lambda_{observed}$ (Å) |
|---|---|---|---|
| C II 6175.4 | $^2F^o$–$^2G$ | em | 6180.8 |
| C II 6213.9 | $^2D^o$–$^2F$ | em | 6218.0 |
| C II 6250.8 | 2p3d $^2D^o$–2p4p $^2P$ | em | 6255.8 |
| C II 6257.2 | $^2P^o$–$^2D$ | em | 6261.8 |
| C II 6259.6 | $^2P^o$–$^2D$ | em | 6264.1 |
| ? | | em | 6284.7 |
| [O I] 6300.3 | $^3P$–$^1D$ | em | 6305.0 |
| Si II 6347.1 + Mg II 6346.8 | $^2S$–$^2P^o$ + $^2D$–$^2F^o$ | em | 6351.2 |
| [O I] 6363.8 | $^3P$–$^1D$ | em | 6368.3 |
| Si II 6371.0 | $^2S$–$^2P^o$ | em | 6376.2 |
| C II 6384.0 | $^4D$–$^4F^o$ | em | 6388.1 |
| C II 6392.0 | $^2G$–$^2H^o$ | em | 6397.2 |
| ? | | em | 6407.5 |
| C II 6424.7 + Na I 6424.7 | $^2D$–$^2F^o$ | em | 6429.9 |
| C II 6462.0 | $^2F^o$–$^2G$ | em | 6466.7 |
| ? | | em | 6511.6 |
| C II 6522.6 | $^2P^o$–$^2D$ | em | 6526.9 |
| [N II] 6548.0 | $^3P$–$^1D$ | em | 6552.9 |
| H$\alpha$ | | em | 6567.7 |
| C II 6578.1 | $^2S$–$^2P^o$ | P Cyg | 6582.4 |
| C II 6582.9 + [N II] 6583.5 | $^2S$–$^2P^o$ | em | 6588.3 |
| ? | | em | 6603.8 |
| C II 6617.6 | $^2D$–$^2F^o$ | em | 6622.9 |
| ? | | em | 6626.7 |
| C II 6667.7 | $^2G$-$^2H^o$ | em | 6673.3 |
| He I 6678.2 | $^1P^o$–$^1D$ | P Cyg | 6682.7 |
| [S II] 6716.4 | $^4S^o$–$^2D^o$ | em | 6721.9 |
| C II 6723.3 | $^2D$–$^2P^o$ | em | 6729.0 |
| C II 6727.3 | $^4D$–$^4D^o$ | em | 6732.2 |
| [S II] 6730.8 | $^4S^o$–$^2D^o$ | em | 6736.2 |
| C II 6733.6 | $^4D$–$^4D^o$ | em | 6739.0 |
| C II 6738.6 | $^4D$–$^4D^o$ | em | 6744.0 |
| C II 6742.4 | $^4D$–$^4D^o$ | em | 6747.7 |
| C II 6750.5 | $^4D$–$^4D^o$ | em | 6756.3 |
| C II 6755.2 | $^4D$–$^4D^o$ | em | 6760.6 |
| ? | | em | 6767.4 |
| C II 6780.6 | $^4P^o$–$^4D$ | P Cyg | 6785.2 |
| C II 6783.9 | $^4P^o$–$^4D$ | em | 6789.8 |
| C II 6787.2 | $^4P^o$–$^4D$ | em | 6793.0 |

| Line ID | Transition | Type | $\lambda_{observed}$ |
|---|---|---|---|
| (Å) | | | (Å) |
| C II 6791.5 | $^4P^o$–$^4D$ | em | 6797.3 |
| C II 6798.1 | $^4P^o$–$^4D$ | em | 6803.5 |
| C II 6800.7 | $^4P^o$–$^4D$ | em | 6806.4 |
| C II 6812.3 | $^4P^o$–$^4D$ | em | 6817.6 |
| C II 6820.6 | $^2F^o$–$^2G$ | em | 6827.1 |
| C II 6874.5 | $^2G$-$^2H^o$ | em | 6881.0 |
| C II 6883.4 | $^2D$–$^2F^o$ | em | 6889.0 |
| ? | | em | 6911.7 |
| ? | | em | 6915.8 |
| ? | | em | 6927.0 |
| Ne II 6930.4? | | em | 6935.5 |
| ? | | em | 6999.8 |
| C III 7037.3 | $^1P^o$–$^1D$ | em | 7042.8 |
| C II 7046.3 | 2p3p $^4S$–2p3d $^4P^o$ | em | 7051.6 |
| C II 7053.1 | 2p3p $^4S$–2p3d $^4P^o$ | em | 7058.5 |
| He I 7065.2 | $^3P^o$–$^3S$ | em | 7070.9 |
| C II 7101.8 | $^2F^o$–$^2G$ | em | 7108.0 |
| C II 7112.5 | $^4D$–$^4F^o$ | em | 7118.1 |
| C II 7115.6 | $^4D$–$^4F^o$ | em | 7121.0 |
| C II 7119.9 | $^4D$–$^4F^o$ | em | 7125.6 |
| C II 7125.7 | $^4D$–$^4F^o$ | em | 7131.2 |
| C II 7134.1 | $^4D$–$^4F^o$ | em | 7139.2 |
| C II 7159.9 | $^2G$-$^2H^o$ | em | 7165.7 |
| C II 7231.3 | $^2P^o$–$^2D$ | em | 7236.9 |
| C II 7236.4 | $^2P^o$–$^2D$ | em | 7242.3 |
| C II 7267.2 | $^2D$–$^2F^o$ | em | 7273.1 |
| He I 7281.4 | $^1P^o$–$^1S$ | em | 7287.0 |
| [O II] 7320.0 | $^2D^o$–$^2P^o$ | em | 7325.2 |
| [O II] 7329.7 | $^2D^o$–$^2P^o$ | em | 7335.7 |
| C II 7370.0 | $^2D$–$^2P^o$ | em | 7375.8 |
| C III 7433.5 | $^3P^o$–$^3D$ | em | 7439.1 |
| ? | | em | 7465.0 |
| ? | | em | 7472.1 |
| C I 7470.1 | $^3P^o$–$^3D$ | em | 7475.2 |
| C II 7505.3 | $^2P^o$–$^2D$ | em | 7510.8 |
| C II 7508.9 | $^2P^o$–$^2P$ | em | 7515.1 |
| C II 7520 | $^2P^o$–$^2P$ | em | 7525.6 |
| ? | | em | 7529.9 |
| C II 7530.6 | $^2P^o$–$^2P$ | em | 7536.3 |

**Table 1** *(continued)*

| Line ID | Transition | Type | $\lambda_{observed}$ |
| --- | --- | --- | --- |
| (Å) | | | (Å) |
| ? | | em | 7564.5 |
| ? | | em | 7570.7 |
| ? | | em | 7575.0 |
| ? | | em | 7580.0 |
| O I 7771.9 | $^5S^o$–$^5P$ | abs | 7776.6 |
| O I 7775.4 | $^5S^o$–$^5P$ | P Cyg | 7780.7 |
| ? | | em | 7802.1 |
| He I? | | em | 7822.2 |
| C II 7860.5 | $^2D$–$^2F^o$ | em | 7866.8 |
| Mg II 7877.1 | $^2P^o$–$^2D$ | em | 7883.0 |
| Mg II 7896.3 | $^2P^o$–$^2D$ | em | 7902.4 |
| C II 8037.7 | $^4P$–$^4P^o$ | em | 8044.6 |
| C II 8062.1 | $^4P$–$^4P^o$ | em | 8068.5 |
| C II 8076.6 | $^4P$–$^4P^o$ | em | 8082.7 |
| C II 8139.6 | $^2F^o$–$^2G$ | em | 8146.5 |
| C II 8214.5 | $^2G$–$^2H^o$ | em | 8221.0 |
| C II 8234.3 | $^2G$-$^2F^o$ | em | 8240.6 |
| He I 8291.1 + 8292.9 | $^3D$–$^3P^o$ + $^1D$–$^1F^o$ | em | 8299.8 |
| He I 8335.8 | $^1P^o$-1S | em | 8341.9 |
| He I 8340.0 | $^1P^o$–$^1D$ | em | 8345.4 |
| He I 8351.4 | $^1P^o$–$^1D$ | em | 8356.6 |
| He I 8393.4 + 8395.0 | $^1P^o$–$^1D$ | P Cyg | 8399.3 |
| C II 8411.0 | $^2D$–$^2F^o$ | em | 8416.6 |
| C II 8413.7 | $^4F^o$–$^4G$ | em | 8419.5 |
| C II 8424.0 + He I 8423.8 | $^4F^o$–$^4G$+$^3D$–$^3P^o$ | em | 8429.7 |
| O I 8446.4 | $^3S^o$–$^3P$ | em | 8452.8 |
| He I 8581.9 + 8582.6 | $^3D$–$^3F^o$+$^3P^o$–$^3D$ | em | 8589.7 |
| C II 8682.5 | $^2S$–$^2P^o$ | em | 8689.5 |
| C II 8696.7 | $^2S$–$^2P^o$ | em | 8703.7 |
| C II 8794.1 | 2p3p $^2D$–2p3d $^2F^o$ | em | 8801.2 |
| C II 8800.3 | 2p3p $^2D$–2p3d $^2F^o$ | em | 8808.1 |
| C II 8873.3 | $^2D$–$^2F^o$ | em | 8880.2 |
| C II 8912.3 | $^2G$-$^2F^o$ | em | 8919.6 |
| C II 9089.1 | $^2G$–$^2H^o$ | em | 9095.8 |
| He I + ? | $^3P^o$-3P | em | 9101.9 |
| Mg II 9218.2 | $^2S$–$^2P^o$ | em | 9225.2 |
| C II 9224.0 | $^2F^o$–$^2G$ | em | 9230.6 |
| C II 9229.6 | $^2D$–$^2F^o$ | em | 9236.9 |
| C II 9236.8 | $^2P^o$–$^2S$ | em | 9243.7 |

**Table 1** *(continued)*

| Line ID (Å) | Transition | Type | $\lambda_{observed}$ (Å) |
|---|---|---|---|
| Mg II 9244.3 | $^2S$–$^2P^o$ | em | 9251.3 |
| C II 9251.0 | $^2P^o$–$^2D$ | em | 9257.3 |